\documentclass[aps,prc,twocolumn,floatfix,superscriptaddress,amsmath,amssymb,nofootinbib]{revtex4-2}

\usepackage{dcolumn}
\usepackage{braket}
\usepackage{float}
\usepackage{graphicx}
\usepackage[colorlinks=true,linkcolor=blue,citecolor=blue,urlcolor=blue]{hyperref}
\usepackage{orcidlink}

\graphicspath{{.}{./figures/}}

\begin{document}

\title{Computing nuclear response functions with time-dependent coupled-cluster theory} 

\thanks{This manuscript has been authored in part by UT-Battelle, LLC, under contract DE-AC05-00OR22725 with the US Department of Energy (DOE). The publisher acknowledges the US government license to provide public access under the DOE Public Access Plan (http://energy.gov/downloads/doe-public-access-plan). This work was also performed under the auspices of the U.S. Department of Energy by Lawrence Livermore National Laboratory under Contract DE-AC52-07NA27344. LLNL-JRNL-2012532.}

\author{Francesca~Bonaiti\orcidlink{0000-0002-3926-1609}}
\affiliation{Facility for Rare Isotope Beams, Michigan State University, East Lansing, Michigan 48824, USA}
\affiliation{Physics Division, Oak Ridge National Laboratory, Oak Ridge, Tennessee 37831, USA}

\author{Cody Balos\orcidlink{0000-0001-9138-0720}}
\affiliation{Center for Applied Scientific Computing, Lawrence Livermore National Laboratory, Livermore, California, 94550, USA}

\author{Kyle Godbey\orcidlink{0000-0003-0622-3646}}
\affiliation{Facility for Rare Isotope Beams, Michigan State University, East Lansing, Michigan 48824, USA}

\author{Gaute~Hagen\orcidlink{0000-0001-6019-1687}}
\affiliation{Physics Division, Oak Ridge National Laboratory, Oak Ridge, Tennessee 37831, USA}
\affiliation{Department of Physics and Astronomy, University of Tennessee, Knoxville, Tennessee 37996, USA}

\author{Thomas~Papenbrock\orcidlink{0000-0001-8733-2849}}
\affiliation{Department of Physics and Astronomy, University of Tennessee, Knoxville, Tennessee 37996, USA}
\affiliation{Physics Division, Oak Ridge National Laboratory, Oak Ridge, Tennessee 37831, USA}

\author{Carol S. Woodward\orcidlink{0000-0002-6502-8659}}
\affiliation{Center for Applied Scientific Computing, Lawrence Livermore National Laboratory, Livermore, California, 94550, USA}

\begin{abstract}
We compute nuclear response functions by solving the time-dependent $A$-body Schr\"odinger equation, recording the time-dependent transition moment and extracting spectral information via Fourier transforms. The solution of the time-dependent many-body problem accounts for correlations on top of the mean field by taking advantage of a time-dependent formulation of coupled-cluster theory. As a validation, we focus on electric dipole transitions in $^4$He and $^{16}$O and compare moments of the response function distribution to the results of an equivalent static framework, finding negligible discrepancies. We investigate how proton and neutron densities evolve in time, and we see the traditional picture of soft and giant dipole resonances as collective oscillations of protons and neutrons emerging from our calculations in $^{16}$O and $^{24}$O. This method also allows us to investigate the behavior of the nucleus in the presence of a strong electric field. In that regime, the behavior of the system becomes chaotic. Qualitatively, the spectral information obtained in this limit is in line with previous time-dependent mean-field results.  
\end{abstract}
\maketitle

\section{Introduction}
Accurate predictions of nuclear dynamical processes help to understand the synthesis of chemical elements in astrophysical environments and their abundances. In particular, nuclear fusion rates in stellar cores, neutron capture cross sections and fission fragment distributions are the ingredients entering astrophysical simulations~\cite{wiescher2025,cowan2021,mumpower2016}. Such nuclear inputs are especially needed for nuclei far from stability, for which experimental information is scarce and one needs to rely on theory.

Up to now, time-dependent density functional theory (TDDFT) has been the primary computational framework capable of describing nuclear dynamics across a wide range of nuclei ~\cite{nakatsukasa2016,maruhn2014,simenel2025}. In particular, TDDFT has achieved considerable success in calculating fission barriers and fragment properties~\cite{simenel2014,scamps2018impact,bulgac2021,bulgac2022}, describing collective excitations~\cite{maruhn2005, nakatsukasa2005,umar2005,reinhard2006,goddard2013}, and modeling heavy-ion collisions~\cite{umar2006,simenel2011,godbey2017,godbey2022}. However, these approaches have limitations and do not describe quantum many-body tunneling for energies below the barrier~\cite{negele1982,simenel2012,nakatsukasa2016}. It is expected that many-body correlations beyond mean-field will remedy such shortcomings  and this motivates the development of strategies to incorporate them~\cite{hasegawa2020,desouza2024,qiang2024}.

One promising avenue is to employ ab initio approaches~\cite{hergert2020,Ekstrom:2022yea}, which combine chiral effective field theory (EFT) interactions~\cite{weinberg1990,epelbaum2009,machleidt2011}, rooted in the underlying theory of quantum chromodynamics, with systematically improvable many-body methods~\cite{hagen2014,dickhoff2004,lee2009,hergert2016,stroberg2019}. Significant progress has been made in ab initio nuclear structure calculations, which have now reached $^{208}$Pb~\cite{hu2022,hebeler2023,arthuis2024} and $^{266}$Pb~\cite{bonaiti2025}. However, first-principles descriptions of nuclear reactions are still limited to light nuclei~\cite{quaglioni2008,elhatisari2015,navratil2016,navratil2020}.

Time-dependent coupled-cluster (TDCC) theory can provide a possible pathway forward. TDCC approaches were developed decades ago to describe nuclear collisions~\cite{hoodbhoy1978,hoodbhoy1979} and spectral functions~\cite{schgun78}. The connection between dynamics and response functions was made even earlier~\cite{arponen1983,dalmon83}. The bivariational approach by  \textcite{kvaal2012} clarified how TDCC should be understood and implemented. In recent years, the method has become popular again, see  Ref.~\cite{sverdrupofstad2020} for a review. The nuclear physics application~\cite{pigg2012} was also based on Kvaal's bivariational approach. That work, however, was limited to a spherically symmetric setting because of available computational resources. Since then computers have advanced so much that it is now feasible to push this method toward more realistic applications in medium-mass nuclei. While we keep axial symmetry in this work, implementations without any symmetry requirements are on the horizon. 

In this work we focus on electromagnetic response functions. This choice is strategic: electromagnetic responses can also be computed within static frameworks~\cite{baccapastore2014,raimondi2019,simonis2019,beaujeault2023,porro2024, burrows2025}, allowing us to easily benchmark the approach. To address the calculation of nuclear responses in TDCC, we apply first-order time-dependent perturbation theory. There, one can relate nuclear response functions to the Fourier transform of the time evolution of the transition moment of interest. 

This paper is organized as follows. In Section~\ref{sec: pert-theory}, we present how to calculate nuclear response functions in a time-dependent approach. In Section~\ref{sec: comp-setup}, we provide details about the nuclear Hamiltonian, the TDCC formalism, the time integration solver, and the main parameters governing our simulations. In Section~\ref{sec: results} we validate our method on nuclear dipole response properties by comparing TDCC results against equivalent static calculations. Moreover, we present two applications where a time-dependent approach offers crucial advantages over a static treatment: the analysis of nuclear density fluctuations and of the behavior of the nucleus when immersed in strong electromagnetic fields. In Section~\ref{sec: conclusion}, we draw our conclusions and give a brief outlook.

\section{Nuclear responses in a time-dependent framework}
\label{sec: pert-theory}
We are interested in computing the nuclear response function
\begin{equation}
    R(E) = \sum_n |\braket{\Psi_n|\Theta|\Psi_0}|^2 \delta(E - E_n + E_0),
    \label{response}
\end{equation}
which describes the behavior of the nucleus in the presence of a small perturbation, induced by a transition operator $\Theta$. In Eq.~(\ref{response}), $\ket{\Psi_0}$ is the ground state of the nucleus, with energy $E_0$, while $\ket{\Psi_n}$ are the excited states with energies $E_n$ that the nucleus can access via the action of $\Theta$. 

Computing Eq.~(\ref{response}) requires the knowledge of the nuclear spectrum, including both bound and unbound excited states. While the starting point for such a calculation is typically the stationary Schrödinger equation, a time-dependent approach can also be used. In this case one can show that the Fourier transform of the expectation value of the transition operator in time is directly related to the response function. This connection was already made in the past, see e.g. Refs.~\cite{arponen1983,dalmon83,umar2005}. In this Section, we summarize the main aspects of the derivation. Let us begin by considering the real-time evolution of the system according to the time-dependent Schrödinger equation
\begin{equation}
    i\hbar \frac{d}{dt} \ket{\Psi(t)} = H(t)\ket{\Psi(t)}.
    \label{tdse}
\end{equation}
Here, $\ket{\Psi(t)}$ is the nuclear wave function in time and the Hamiltonian $H(t)$ is defined as
\begin{equation}
    H(t) = H_0 + \varepsilon W(t).
    \label{hamiltonian}
\end{equation}
where $H_0$ is the intrinsic nuclear Hamiltonian, and 
\begin{equation}
    W(t) = \Theta \delta(t)
\end{equation}
introduces a pulse at the initial time $t=0$, with strength $\varepsilon$. At $t=0$, we assume that the system is in its ground state
\begin{equation}
    \ket{\Psi(t = 0)} = \ket{\Psi_0}. 
    \label{incon}
\end{equation}

For small values of $\varepsilon$, we can find a solution for $\ket{\Psi(t)}$ in Eq.~(\ref{tdse}) using first-order time-dependent perturbation theory~\cite{cohen1977}. 
In this case the solution of Eq.~(\ref{tdse}) reads
\begin{equation}
    \begin{split}
        \ket{\Psi(t)} &= 
        e^{-iE_0 t/\hbar} \ket{\Psi_0}\\&-\frac{i\varepsilon}{2\hbar} \sum_n \braket{\Psi_n|\Theta|\Psi_0} e^{-iE_n t/\hbar} \ket{\Psi_n} + \mathcal{O}(\varepsilon^2).
    \end{split}
    \label{psit}
\end{equation}
where $\ket{\Psi_0}$ and $\ket{\Psi_n}$ are the ground state and excited states of $H_0$, with energy $E_0$ and $E_n$, respectively. Using Eq.~(\ref{psit}) one can compute the time-dependent transition moment, i.e. the expectation value of the transition operator in time $\Theta(t) = \braket{\Psi(t)|\Theta|\Psi(t)}$. Retaining only linear terms in $\varepsilon$ in Eq.~(\ref{psit}), we obtain
\begin{equation}
\begin{split}
    \Theta(t) 
     &= - \frac{\varepsilon}{\hbar} \sum_n |\braket{\Psi_n|\Theta|\Psi_0}|^2 \sin(\omega_{n0}t) + O(\varepsilon^2) \\
\end{split}
\label{dt}
\end{equation}
with $\omega_{n0} = (E_n - E_0)/\hbar$. By calculating the Fourier transform $\widetilde{\Theta}(\omega)$ of Eq.~(\ref{dt}), where $\omega$ is the angular frequency, we obtain
\begin{equation}
\begin{split}
    \widetilde{\Theta}(\omega) &= - 
     \frac{i\pi\varepsilon}{\hbar} \sum_n |\braket{\Psi_n|\Theta|\Psi_0}|^2 [\delta(\omega - \omega_{n0}) -  \delta(\omega + \omega_{n0})]. \\
\end{split}
\end{equation}
Focusing on physical solutions with $\omega_{n0} > 0$, and rewriting the transform in terms of the excitation energy $E$, we get
\begin{equation}
    \begin{split}
    \widetilde{\Theta}(E) 
    &= -i\pi\varepsilon \sum_n |\braket{\Psi_n|\Theta|\Psi_0}|^2 \delta(E - E_n + E_0) \\&= -i\pi\varepsilon R(E),
    \end{split}
\end{equation}
where we recognize the response function $R(E)$ as
\begin{equation}
    R(E) = \mathrm{Im}\left(\frac{\widetilde{\Theta}(E)}{\pi\varepsilon}\right).
    \label{tdresponse-delta}
\end{equation}

Two comments are in order. First, in this derivation we focused on the case of a perturbation with a time profile given by a Dirac delta pulse. As shown in Ref.~\cite{calvayrac1997}, a similar expression can be recovered in the case of a generic time profile $g(t)$
\begin{equation}
    R(E) = \mathrm{Im}\left(\frac{\widetilde{\Theta}(E)}{\varepsilon \widetilde{g}(E)}\right).
    \label{tdresponse}
\end{equation}
where $\widetilde{g}(E)$ is the Fourier transform of $g(t)$. Second, nuclear response functions can be employed to describe the behavior of the system only in the perturbative limit, where $\varepsilon$ is small. In general, time-dependent methods allow for the exploration of a larger range of phenomena by varying the strength parameter $\varepsilon$, as for instance non-linear or even chaotic effects~\cite{reinhard2007,vretenar1997}. In the case of nuclear response functions, static and time-dependent approaches will match only when the chosen value of $\varepsilon$ falls within the linear perturbative limit. We will explore the non-linear regime in Section~\ref{sec: nonlinear}.

\section{Computational setup}
\label{sec: comp-setup}
To compute nuclear response functions in a time-dependent framework, we need to solve Eq.~(\ref{tdse}) and follow the evolution of the transition moment $\Theta(t)$. This goal can be achieved by extending the coupled-cluster ansatz typically employed in solving the stationary Schrödinger equation to the time-dependent domain.   In this Section we discuss details of the Hamiltonian $H(t)$ employed in our computations, review the relevant aspects of TDCC from Ref.~\cite{pigg2012}, and describe the ordinary differential equation solver employed in this work. Details on the parameters governing TDCC simulations are also provided. 

\subsection{Time-dependent Hamiltonian}
The starting point of our computations is the intrinsic nuclear Hamiltonian
\begin{equation}
        H_0 = \sum_{i<j} \left(\frac{(\mathbf{p}_i -\mathbf{p}_j)^2}{2mA} + V^{\rm NN}_{ij}\right). 
        \label{H0}
\end{equation}
which is given by the intrinsic kinetic energy (i.e. the kinetic energy of the center of mass is removed) and by the
nucleon-nucleon interaction. In all the calculations presented in this work, we employ the chiral nucleon-nucleon interaction NNLO$_{\rm opt}$~\cite{ekstrom2013}. The inclusion of three-nucleon forces is left for the future. 

The nature of the response function computed via Eq.~(\ref{response}) depends on the perturbation operator $\Theta$ entering Eq.~(\ref{hamiltonian}). Throughout this work, we focus on the case of electric dipole transitions, driven by the operator
\begin{equation}
    D = \sum_{i = 1}^A (\mathbf{r}_i - \mathbf{R}_{\rm CM}) \left(\frac{1+\tau_{i,z}}{2}\right) \ , 
\end{equation}
where $\mathbf{r}_i$ and $\mathbf{R}_{\rm CM}$ are the position of the $i$-th nucleon and of the center of mass, respectively, while $\tau_{i,z}$ is the projection of the isospin operator on the $z$-axis. 

As the electric dipole operator is not a scalar, the time-dependent Hamiltonian~(\ref{hamiltonian}) conserves neither angular momentum nor parity. Therefore, time evolution breaks those symmetries, and when considering a single-particle basis, only the $z$-axis projections of the angular momentum $j_z$ and of isospin $t_z$ of the orbitals will be preserved.  

For response function calculations, we use a perturbation strength $\varepsilon = 0.1$ MeV/fm, which, as shown later in Section~\ref{sec: nonlinear}, turns out to be well within the linear perturbative regime. The dependence on time of Eq.~(\ref{hamiltonian}) is set by the time profile $g(t)$. Instead of a Dirac delta pulse we employ a Gaussian peaking at a time $t_0$
\begin{equation}
    g(t) = \frac{1}{\sqrt{\pi}} e^{-\left({t-t_0\over \tau_{\rm wid}}\right)^2}.
    \label{time-profile}
\end{equation}
In our calculations, we set $t_0 = 3$~fm/$c$ and $\tau_{\rm wid}=1$~fm/$c$. As $t_0\gg \tau_{\rm wid}$ holds approximately, the perturbation is vanishingly small at $t=0$ fm/$c$ and beyond $t\approx 2t_0$.  

\subsection{Time-dependent coupled-cluster theory}
In TDCC theory, we write the nuclear wave function according to the ansatz
\begin{equation}
    \ket{\Psi(t)} = e^{T(t)} \ket{\Phi_0},
    \label{tdcc-ansatz}
\end{equation}
where 
\begin{equation}
    \ket{\Phi_0} = \prod_{i=1}^{A} a^{\dagger}_i(t)\ket{0}
    \label{ref}
\end{equation}
is an $A$-nucleon product state, acting as reference state. Here, $a^{\dagger}_p$ creates a nucleon in state $|p\rangle = a^{\dagger}_p|0\rangle$, and $a_p$ is the corresponding annihilation operator. While in general the single-particle wave functions are allowed to depend on time, here we consider the case of a static reference. From Eq.~(\ref{ref}) on we have already used the convention that indices $i,j,k,\dots$ label states occupied in the reference state, while $a,b,c,\dots$ label unoccupied states. Indices $p,q,\dots$ will be used to label arbitrary single-particle states. 

Equation~(\ref{tdcc-ansatz}) also depend on the cluster operator $T(t)$, which can be expanded as
\begin{equation}
    T(t) = t_0(t) + T_1(t) + T_2(t) + T_3(t) + \dots,
\end{equation}
Here $t_0(t)$ is a complex phase, while $T_1(t)$, $T_2(t)$, $T_3(t)$, etc.\;introduce 1p-1h, 2p-2h, 3p-3h and up to $A$p-$A$h excitations on top of $\ket{\Phi_0}$
\begin{equation}
    T_n(t) = {1\over (n!)^2}\sum_{\substack{i_1,\dots,i_n \\a_1,\dots,a_n}} t^{a_1,\dots,a_n}_{i_1,\dots,i_n}(t) a^{\dagger}_{a_1} \dots a^{\dagger}_{a_n} a_{i_n} \dots a_{i_1} .
\end{equation}
The amplitudes $t^{a_1,\dots,a_n}_{i_1,\dots,i_n}(t)$ depend on time. In this work, we truncate this expansion at the CCSD level
\begin{equation}
    T(t) = t_0(t) + \sum_{ia} t^a_i(t) a^{\dagger}_a a_i + {1\over 4}\sum_{ijab} t^{ab}_{ij}(t) a^{\dagger}_a a^{\dagger}_b a_j a_i.
\end{equation}

Our goal is now to determine the real-time evolution of $t_0(t)$ and of the singles and doubles amplitudes. Substituting the TDCC ansatz of Eq.~(\ref{tdcc-ansatz}) in Eq.~(\ref{tdse}), we obtain the time-dependent Schrödinger equation in the coupled-cluster formalism  
\begin{equation}
    i\hbar e^{-T(t)} \partial_t e^{T(t)} \ket{\Phi_0} = \overline{H} \ket{\Phi_0},
    \label{tdsecc}
\end{equation}
where
\begin{equation}
    \overline{H} = e^{-T(t)} H(t) e^{T(t)}
\end{equation}
is the similarity-transformed Hamiltonian. Using the Baker-Campbell-Hausdorff expansion, the product $e^{-T(t)} \partial_t e^{T(t)}$ can be rewritten as
\begin{equation}
    e^{-T(t)} \partial_t e^{T(t)} = \partial_t + \dot{T}(t) + \frac{1}{2} [\dot{T}(t), T(t)] +\dots \ . 
\end{equation}
If the single-particle wavefunctions do not depend on time, the above equation reduces to:
\begin{equation}
    e^{-T(t)} \partial_t e^{T(t)} = \partial_t + \dot{T}(t).
    \label{derivative}
\end{equation}
Using Eq.~(\ref{derivative}) in Eq.~(\ref{tdsecc}) and multiplying on the left by the adjoints of the states $\ket{\Phi_0}$, $\ket{\Phi^a_i} \equiv a^{\dagger}_a a_i \ket{0}$, and $\ket{\Phi^{ab}_{ij}} \equiv  a^{\dagger}_a a^{\dagger}_b a_j a_i \ket{0}$ we obtain the time-dependent equations for the coupled-cluster amplitudes at the CCSD level
\begin{equation}
\begin{split}
    i\hbar \dot{t}_0(t) &= \braket{\Phi_0|\overline{H}|\Phi_0}, \\
    i\hbar \dot{t}_i^a(t) &= \braket{\Phi^a_i|\overline{H}|\Phi_0}, \\
    i\hbar \dot{t}_{ij}^{ab}(t) &= \braket{\Phi^{ab}_{ij}|\overline{H}|\Phi_0}.
\end{split}
\label{ampeqtdcc}
\end{equation}  
This is a closed system of ordinary differential equations that must be solved numerically, and we deal with an initial value problem. At $t=0$, we start with the CCSD amplitudes that solve the Hamiltonian~(\ref{H0}); this yields the energy $E_0$.

Since the similarity-transformed Hamiltonian is non-Hermitian, calculating expectation values such as the time-dependent transition moment requires also the knowledge of the time-dependent left state $\bra{\widetilde{\Psi}(t)}$. This is the key of the bivariational approach~\cite{kvaal2012}. For the left state one uses the ansatz
\begin{equation}
    \bra{\widetilde{\Psi}(t)} = \bra{\Phi_0} L(t) e^{-T(t)} \ ,
    \label{lambdaansatz}
\end{equation}
where $L(t)$ is a de-excitation operator
\begin{equation} 
    L(t) = l_0(t) + \sum_{ia} l^a_i(t) a^{\dagger}_i a_a + {1\over 4}\sum_{ijab}  l^{ab}_{ij}(t) a^{\dagger}_i a^{\dagger}_j a_b a_a + \dots.
\end{equation}
For consistency with $T(t)$, we also truncate $L(t)$ at the CCSD level. Inserting Eq.~(\ref{lambdaansatz}) in the time-dependent Schrödinger equation, we obtain
\begin{equation}
\begin{split}
    -i\hbar \dot{l}_0(t) &= 0, \\
    -i\hbar \dot{l}_i^a(t) &= \braket{\Phi_0|L(t)\overline{H}|\Phi^a_i}, \\
    -i\hbar \dot{l}_{ij}^{ab}(t) &= \braket{\Phi_0|L(t)\overline{H}|\Phi^{ab}_{ij}}. 
    \label{lambdaeqnstdcc}
\end{split}
\end{equation}
This is another set of ordinary differential equations.
We see that the time-dependent cluster amplitudes $T(t)$ do enter the time evolution of $L(t)$, but the latter do not enter the evolution of the former, see Eq.~(\ref{ampeqtdcc}). The initial values $L(0)$ are obtained from solving the left eigenvalue problem
\begin{equation}
    \langle \Psi_0|L(0)\overline{H_0} = E_0 \langle \Psi_0|L(0) \ .
\end{equation}

Once the time evolution of the $T(t)$ and $L(t)$ amplitudes is known, we can compute the one-body density matrix at time $t$
\begin{equation}
    \rho_{pq}(t) = \braket{\phi_0|L(t)e^{-T(t)} a^{\dagger}_p a_q e^{T(t)}|\Phi_0}. 
    \label{eq:density-mat}
\end{equation}
The latter can then be employed to calculate the expectation value of any one-body operator. We are particularly interested in the time-dependent electric dipole moment
\begin{equation}
    D(t) = \braket{\widetilde{\Psi}(t)|D|\Psi(t)} = \operatorname{Tr} D\rho(t).
    \label{trace}
\end{equation}

A comment is in order. The expectation value in Eq.~(\ref{trace}) is real. However, an imaginary component can arise in Eq.~(\ref{trace}) as the $T(t)$ and $L(t)$ amplitudes are truncated. Following previous works in quantum chemistry~\cite{pedersen1997}, we address this issue by only calculating the Fourier transform of the real part of Eq.~(\ref{trace}) in our computations of the response~(\ref{tdresponse}). 

\subsection{Solving the time-dependent coupled-cluster equations}
\label{sec: cvode-solvers}
The ordinary differential equations for the amplitudes of the coupled-cluster operator, $T(t)$, and de-excitation operator, $L(t)$, are computationally expensive to solve.  They become increasingly stiff with both increasing mass number and increasing size of the employed single-particle basis. Thus, an efficient solver is needed.
By rearranging Eq.~\eqref{ampeqtdcc} and Eq.~\eqref{lambdaeqnstdcc} to the form
\begin{equation}
    \dot{y} = f(t,y),~ y(t_0) = y_0,
\label{explicit_form_ode}
\end{equation}
where $y=({t}_0, {t}_i^a, {t}_{ij}^{ab}, {l}_0, {l}_i^a,  {l}_{ij}^{ab})$ is the vector of amplitudes, we can employ the adaptive implicit time integration methods available in the CVODE package of the SUNDIALS library \cite{hindmarsh2005sundials,gardner2022sundials} with only a few modifications to allow for our complex-valued states. The modifications primarily involved development of an implementation of the SUNDIALS vector data structure that works on complex-valued data and some other minor changes to arithmetic within the iterative algebraic solvers. 

For TDCC simulations involving nuclei with small mass numbers (such as $^4$He) and small single-particle basis sizes (including, for example, a maximum number of major oscillator shells $N_{\rm max} = 4$), Eq.~\eqref{explicit_form_ode} is relatively nonstiff. In these cases, we employ the Adams-Moulton methods in CVODE, combined with the SUNDIALS fixed-point iterative nonlinear solver, to efficiently solve the implicit system. 
For TDCC simulations of nuclei with larger mass numbers and larger basis sizes, Eq.~\eqref{explicit_form_ode} is stiff. Here, we utilize the BDF methods in CVODE, paired with either the SUNDIALS modified Newton’s method solver or, in moderately stiff scenarios, the SUNDIALS Anderson-accelerated fixed-point iterative solver \cite{anderson1965iterative,walker2011anderson}. When using the modified Newton approach, the resulting linear system at each Newton step is solved using a scaled, unpreconditioned, GMRES Krylov method. The necessary Jacobian-vector product, $\partial_y f v$, is approximated within SUNDIALS via finite differences, eliminating the need to explicitly form or store the Jacobian.

\begin{figure}[htb]
    \centering
    \includegraphics[width=0.49\textwidth]{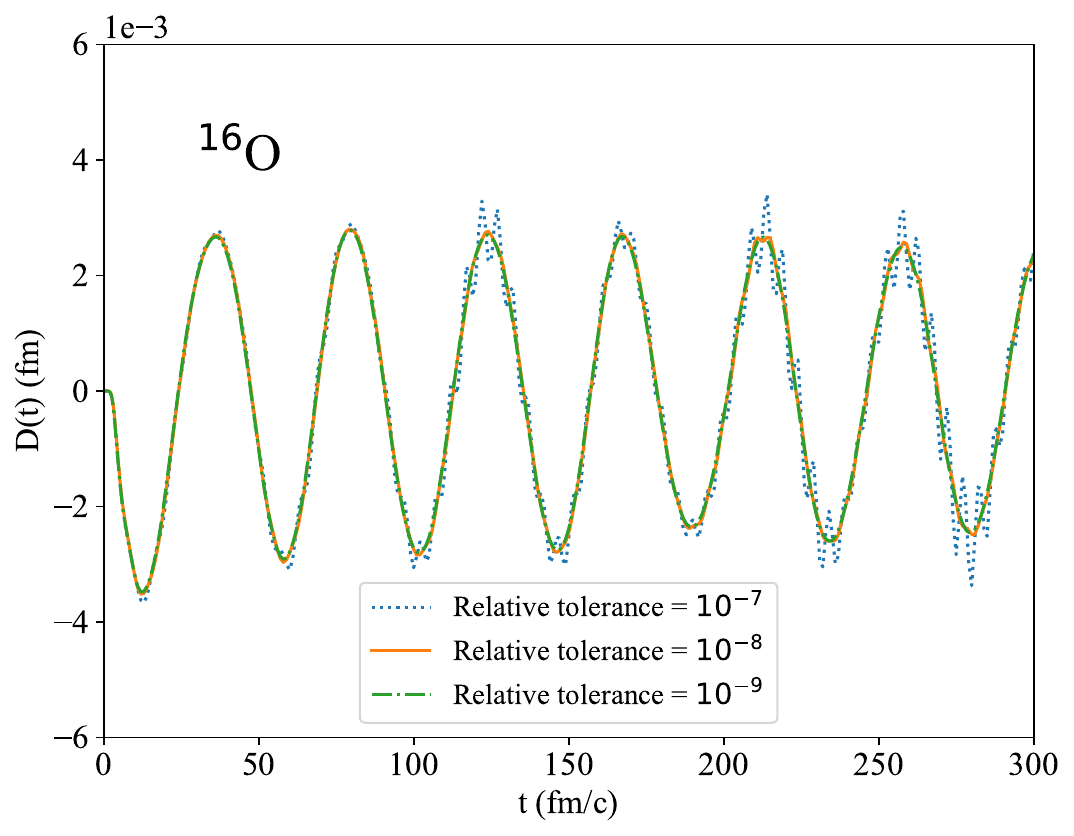}
    \caption{Convergence of the dipole moment with CVODE's relative tolerance size for $^{16}$O and $N_{\rm max} = 6$.}
    \label{fig:tolerance-study-dipole}
\end{figure}

CVODE adapts the time step size used during integration based on an internal error estimate and user-provided relative and absolute error tolerances.
It is important to note that we instruct CVODE to return the solution to Eq.~\eqref{explicit_form_ode} at an interval of $\Delta t$, but internally CVODE may use time steps smaller, or even larger than $\Delta t$. 
Selecting appropriate tolerances is crucial for an efficient solution: too small tolerances will result in unnecessarily small CVODE time steps, and computations will be slow, while too large  tolerances will result in unacceptable error in the solution. 
Setting the absolute tolerances is more straightforward, as they represent component-specific noise levels. As such, the absolute tolerances can be chosen to be the smallest magnitudes of the amplitudes that we care to resolve. Throughout all of the computations in this paper, we have set them to $10^{-9}$. The relative tolerance controls the relative error, that is, a relative tolerance of $10^{-4}$ will control errors to 0.01\%. Figure~\ref{fig:tolerance-study-dipole} shows three TDCC simulations of the time-dependent dipole moment of $^{16}$O with relative tolerances of $10^{-7}, 10^{-8}$, and $10^{-9}$. Based on these results, we see that a relative tolerance of $10^{-8}$ is sufficient.

Since CVODE may choose time steps larger than $\Delta t$ internally, we have found that it is useful, from an efficiency perspective, to set a cap of $\Delta t$ on the CVODE step size (an option that can be controlled) for the duration of the perturbation, set by the time interval where the pulse $g(t)$ is significantly different from $0$.
After this period the cap is removed. Without applying this cap, the tolerances must be tightened to prevent CVODE from stepping over the more rapid dynamics present during the initial transient perturbation.

\subsection{From time to energy domain}
\label{sec:finite-sim-time}
TDCC simulations are characterized by three parameters: the time step $\Delta t$ at which we record the solution of Eqs.~(\ref{ampeqtdcc}) and (\ref{lambdaeqnstdcc}), the number of time samples $N_{\rm samples}$ and the total simulation time $T_{\rm max} = N_{\rm samples}\Delta t$.

The inverse time step $(\Delta t)^{-1}$ defines the highest excitation energy that one can resolve in the Fourier domain. In our computations, we use different time steps as the simulation proceeds. For $0 \leq t \leq 2t_0=6$~fm/$c$, where the Gaussian time profile of Eq.~(\ref{time-profile}) varies significantly, we adopt $\Delta t = 0.2$~fm/$c$. As $\Delta t \ll \tau_{\rm wid}$,  this allows us to accurately sample the time profile of the perturbation while evolving the TDCC equations. Since for the nuclei considered in this work the fastest oscillations in the spectrum have a period larger than $20$~fm/$c$, we can safely increase $\Delta t$ to $1$~fm/$c$ for $t > 6$~fm/$c$. This time step still resolves excitation energies exceeding $200$~MeV. 

The total simulation time $T_{\rm max}$ determines the spectral resolution achievable in the Fourier domain. The smallest energy scale that can be resolved is
\begin{equation}
    \Delta E = \frac{2\pi\hbar}{T_{\rm max}}.
    \label{resolution}
\end{equation}
For example, for $T_{\rm max} = 2000$~fm/$c$, we obtain a resolution of $\Delta E \approx 0.6$ MeV. It becomes clear that $T_{\rm max}$ is a crucial factor in determining the computational cost of TDCC calculations.

Once $D(t)$ is available, the response is obtained from the discrete Fourier transform $\widetilde{D}(E)$ of the finite time-dependent signal. Additional post-processing analysis is also needed in this case. The finite simulation time $T_{\rm max}$ affects the shape of $\widetilde{D}(E)$ by introducing artifacts. This shortcoming can be reduced by truncating the time signal more softly, i.e. by multiplying $D(t)$ with a so-called windowing function before applying the discrete Fourier transform algorithm. An appropriate choice of the windowing function ensures a smooth decay to $0$ of the time signal within the maximum simulation time while slightly reducing the spectral resolution. Following earlier works in TDDFT~\cite{reinhard2006,maruhn2014}, we employ the windowing function
\begin{equation}
    \left[\cos\left(\frac{\pi t}{2T_{\rm max}}\right)\right]^{N_{\rm filt}}
\end{equation}
with even cut-off powers $N_{\rm filt}$. In this work, we employ $N_{\rm filt} = 4$. We have verified that the choice of $N_{\rm filt}$ has a negligible effect on the final spectrum and on TDCC predictions of sum rules.

\section{Results}
\label{sec: results}
In this Section, we present results for the nuclear dipole response and dipole-induced density fluctuations of $^4$He, $^{16}$O, and $^{24}$O. We benchmark our time-dependent results comparing them to those of an equivalent static framework in $^4${He} and $^{16}$O, and discuss the convergence of our results with respect to the model space size and the maximum simulation time. We consider how proton and neutron densities oscillate in time for $^{16}$O and $^{24}$O. In the end, we analyze the behavior of the nucleus in the non-linear regime limit ($\varepsilon \gg 1$ MeV/fm), focusing on the case of $^{16}$O.

In our TDCC calculations, we start from a Hartree-Fock single-particle basis. Model space convergence is controlled by the maximum number of major oscillator shells $N_{\rm max}$ included in the computation, which we vary between $N_{\rm max} = 4$ and $N_{\rm max} = 8$, and by the harmonic oscillator frequency $\hbar\Omega$. For $^{4}$He and $^{16}$O we employ $\hbar\Omega = 16$ MeV, while for $^{24}$O we adopt a lower value of $\hbar\Omega = 12$ MeV, which has been found to have a good convergence behaviour for this nucleus~\cite{bonaiti2024}. The results presented are obtained with $T_{\rm max} = 2000$~fm/$c$. Convergence with respect to $T_{\rm max}$ is discussed in Section~\ref{sec: tmax-conv}. As for static calculations, we employ a spherical Hartree-Fock single-particle basis.

\subsection{Time-dependent dipole moment}
Let us first recall a few relevant scales. Typical dipole transition strength between discrete states in nuclei are of the size $[B(E1)]^{1/2}\approx {\cal O}(0.01 - 0.1$~e\;$ \mathrm{fm})$, see, e.g., Refs.~\cite{tilley1993,summers2007,horiuchi2012,spieker2015}. Thus, linear response theory should use strengths that are perturbatively small compared to these values. Photon energies in nuclear transitions are of the order of ${\cal O}(1-10~\mathrm{MeV})$, and the corresponding wave lengths are much larger than nuclear radii.  

\begin{figure}[htb]
    \centering
    \includegraphics[width=0.49\textwidth]{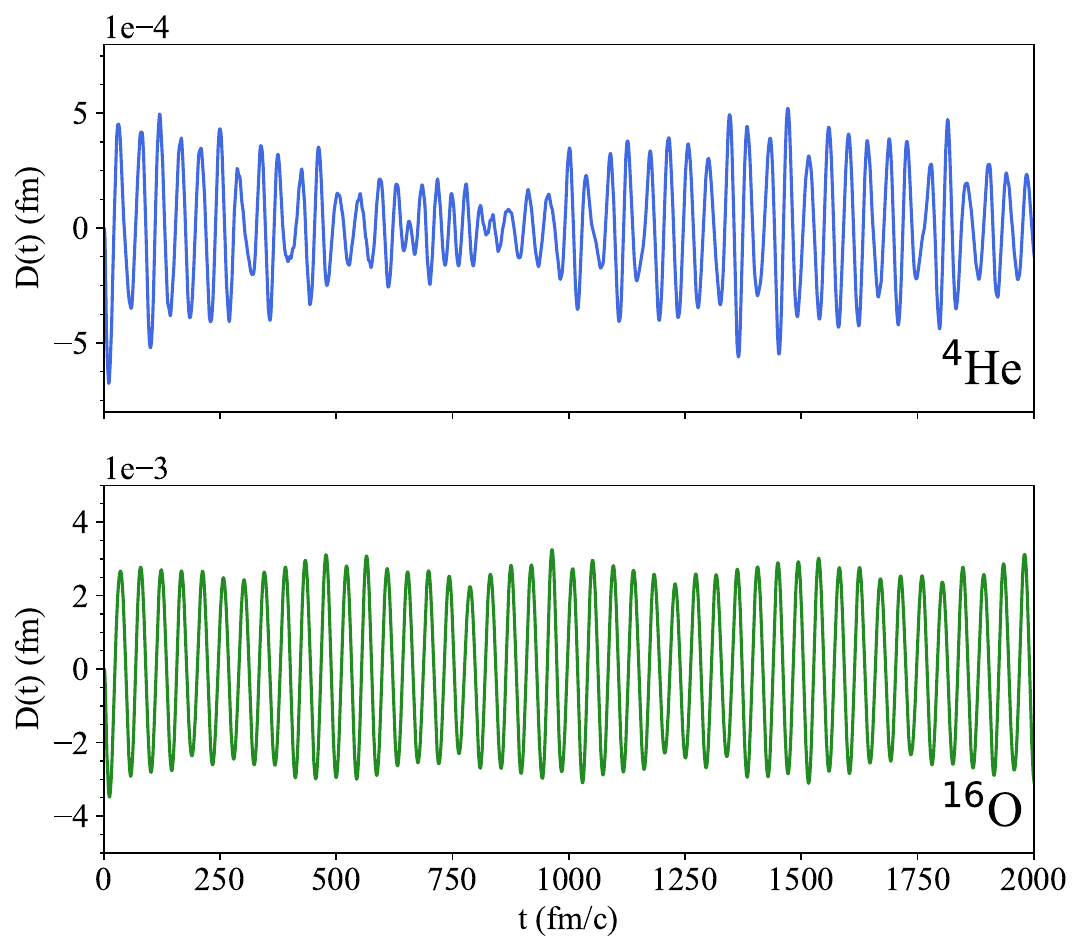}
    \caption{Time-dependent dipole moment for $^{4}$He and $^{16}$O as a function of time for $N_{\rm max} = 6$. }
    \label{fig: redt-conv}
\end{figure}

To start our analysis, it is useful to look at the evolution in time of the real part of the time-dependent dipole moment, from which we can already deduce some spectral information. In Fig.~\ref{fig: redt-conv} we show the dipole moments of $^4$He and $^{16}$O. 

We observe that the time-dependent signals are characterized by a periodic behavior, with almost no damping. In both cases, we can identify a main frequency mode with a period of around $40$~fm/$c$, which correspond to an energy of about $30$~MeV. This is close to the typical energies of the giant dipole resonance (GDR). 

\subsection{Validation in $^4${He} and $^{16}$O}
In this Section, we validate the approach described in Sections~\ref{sec: pert-theory} and~\ref{sec: comp-setup} by comparing the nuclear dipole response functions of $^{4}$He and $^{16}$O obtained from static and time-dependent calculations.

In the linear regime, where static and time-dependent calculations can be compared, we expect symmetry breaking effects induced by time evolution, leading, e.g., to mixing of other modes in the response, to be small. To quantify them, we focus on $^4$He and $N_{\rm max} = 4$, and track the time evolution of the expectation value of the total angular momentum $\braket{J^2}$. Indeed, we find that $\braket{J^2}$ deviates from $0$ by less than $10^{-6}$ for a field strength of $\varepsilon = 0.1$ MeV/fm, while it increases for larger values of $\varepsilon$, reaching around $10^{-1}$ for $\varepsilon = 50$ MeV/fm. As shown later in Section~\ref{sec: nonlinear}, such a field strength leads to non-linear behavior.

For the static computations, we employ the Lorentz integral transform (LIT)~\cite{efros1994,efros2007} combined with the coupled-cluster method~\cite{bacca2013,bacca2014}. Since the latter is also based on coupled-cluster, it allows for a clean comparison within the same many-body approximation scheme. The LIT avoids the calculation of all the bound and continuum excited states of the Hamiltonian in the response~(\ref{response}) by considering an integral transform with Lorentzian kernel of $R(E)$
\begin{equation}
    L(\sigma, \Gamma) = \int dE\; \frac{R(E)}{(E-\sigma)^2 + \Gamma^2} \ .
    \label{lit}
\end{equation}
Here $\sigma$ is the centroid and $\Gamma$ is the width of the kernel. Computing such a transform requires the solution of a coupled-cluster equation-of-motion with a source term but only includes ground-state properties. Once $L(\sigma, \Gamma)$ is computed, $R(E)$ can be recovered by means of an inversion procedure~\cite{efros2007}.  

\begin{figure}[htb]
    \centering
    \includegraphics[width=0.49\textwidth]{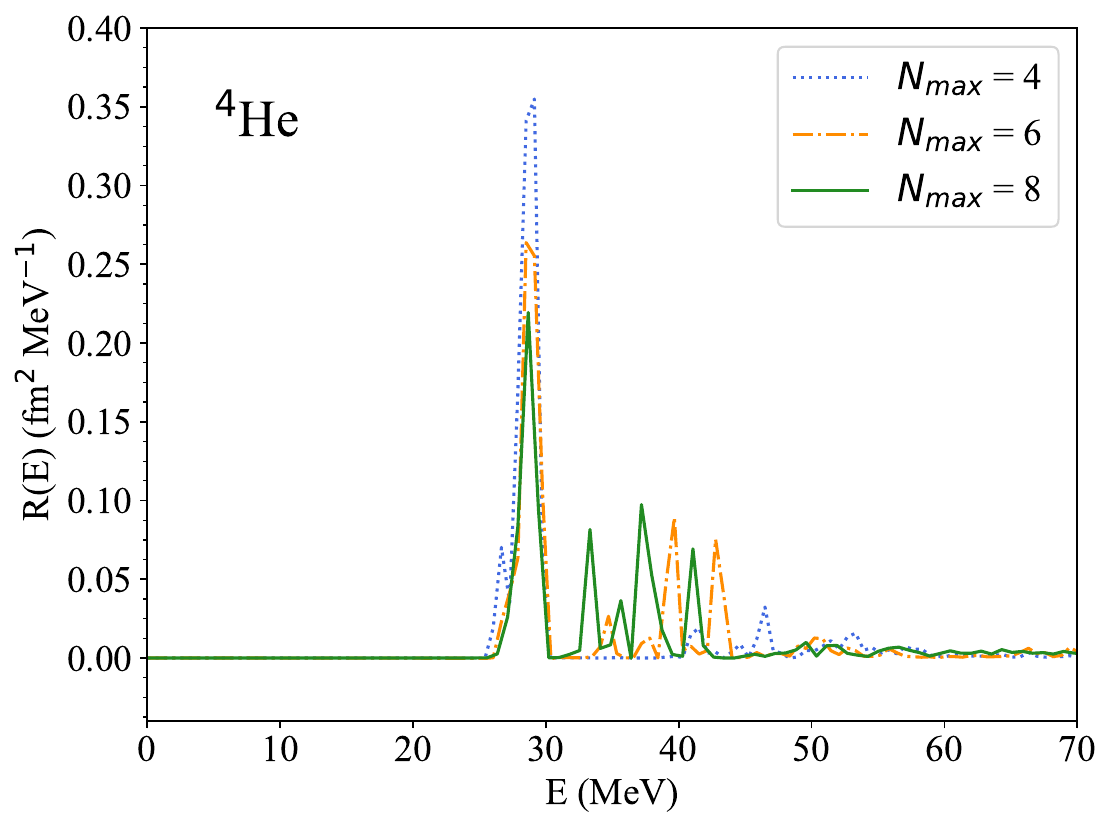}
    \caption{Convergence of the response with model-space size for $^4$He. }
    \label{fig:response-conv-4He}
\end{figure}

\begin{figure}[htb]
    \centering
    \includegraphics[width=0.49\textwidth]{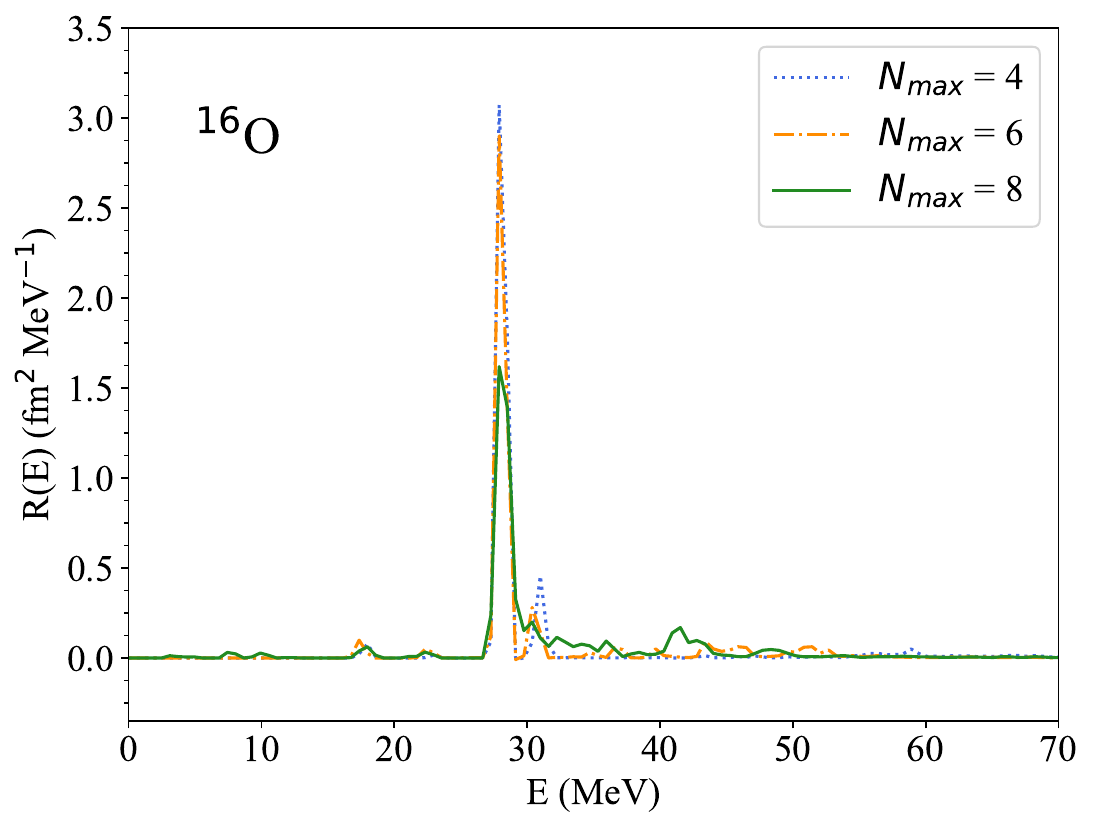}
    \caption{Convergence of the response with model-space size for $^{16}$O. }
    \label{fig:response-conv-16O}
\end{figure}

We start by analyzing the convergence of the TDCC response with respect to the model space size. Results for $^{4}$He and $^{16}$O are shown in Figs.~\ref{fig:response-conv-4He} and~\ref{fig:response-conv-16O}, respectively.  

In the case of $^{4}$He, the position of the main peak of the response, corresponding to the GDR, appears to be converged with respect to the model-space size. Its location, at around $28$ MeV, is consistent with no-core shell model (NCSM) and symmetry-adapted NCSM calculations employing the same interaction, which predict it at $26.5$ MeV~\cite{burrows2025}. The remaining strength is fragmented into smaller peaks between $30$ and $50$ MeV, which tend to move to lower energies with increasing $N_{\rm max}$. 

For $^{16}$O the response is dominated by the GDR peak, at around $28$ MeV. While the $N_{\rm max} = 4$ and $6$ results are almost overlapped, at $N_{\rm max} = 8$ some of the strength in the mean peak is distributed to states at higher energy, between $30$ and $45$ MeV. Nevertheless, the location of the peak remains stable with increasing model-space size. The $N_{\rm max} = 8$ calculation for $^{16}$O is the most time-consuming one presented in this work. Given the maximum wall time allowed by the supercomputing resources at our disposal, $T_{\rm max} = 2000$~fm/$c$ is the maximum simulation time available for this configuration. We leave further optimization of our TDCC machinery to future work. Despite this limitation,  we are able to resolve two low-lying excited states in the response at $8$ and $10$ MeV. The latter are consistent within our resolution with experimental data, indicating the presence of two $1^-$ excited states in the $^{16}$O spectrum at $7.116$ and $9.585$ MeV~\cite{nudat}.

As for benchmark observables, we consider moments of the response function distribution
\begin{equation}
    m_n = \int dE\; E^n R(E) \ .
\end{equation}
Here we focus on the electric dipole polarizability 
$\alpha_D \equiv 2\alpha m_{-1}$ (where $\alpha$ is the fine-structure constant) , the non-energy-weighted sum rule $m_0$, and the energy-weighted sum rule $m_1$. According to Eq.~(\ref{resolution}), TDCC calculations give access to a response function with finite resolution $\Delta E$ set by $T_{\rm max}$. Such a quantity is not directly comparable to the result of a LIT inversion, which in principle is independent of the choice of the Lorentzian width $\Gamma$. However, we can easily determine the LIT of the TDCC response by plugging Eq.~(\ref{tdresponse}) in Eq.~(\ref{lit}). Therefore, in order to provide a test on the strength function distribution we also compare the static and time-dependent LITs for different values of $\Gamma$. 

\begin{figure*}[htb]
    \centering
    \includegraphics[width=0.95\textwidth]{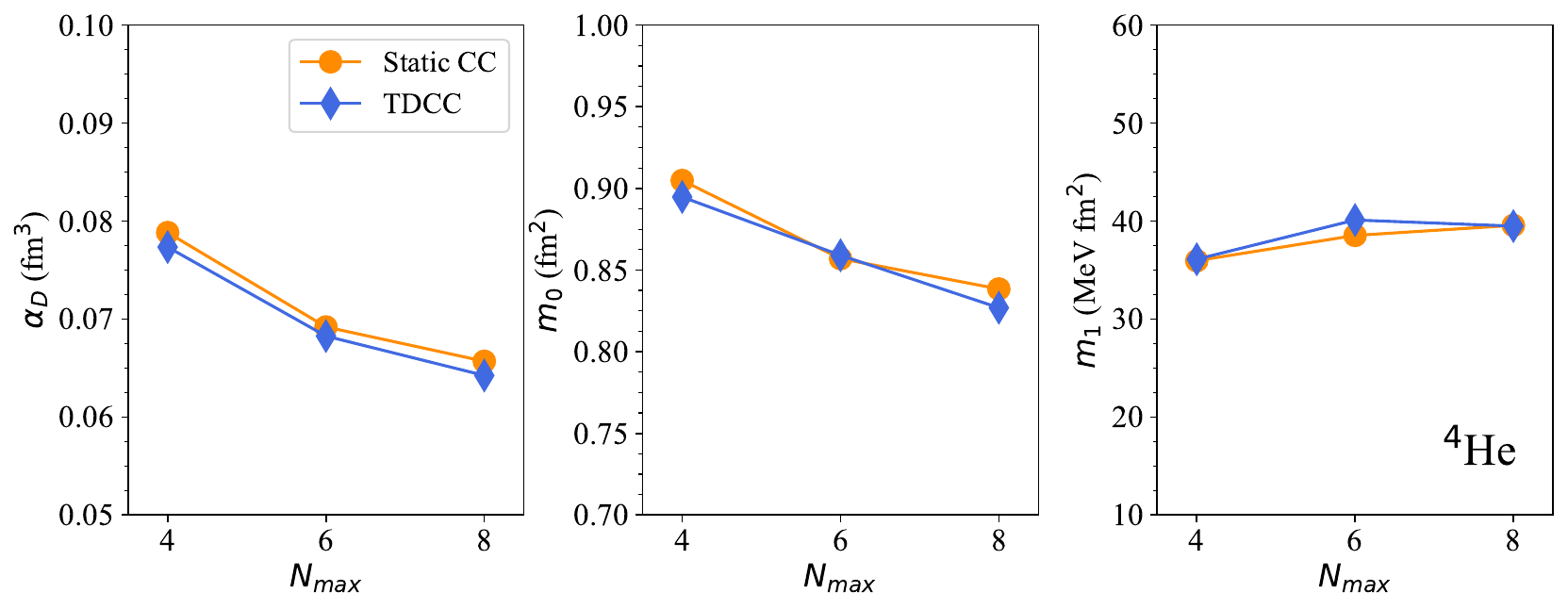}
    \caption{Comparison between the static and time-dependent results for the dipole polarizability $\alpha_D$, the non-energy-weighted sum rule $m_0$ and the energy-weighted sum rule $m_1$ in $^4$He for different model-space sizes. }
    \label{fig: sr-bench-4he}
\end{figure*}

\begin{figure*}[htb]
    \centering
    \includegraphics[width=0.95\textwidth]{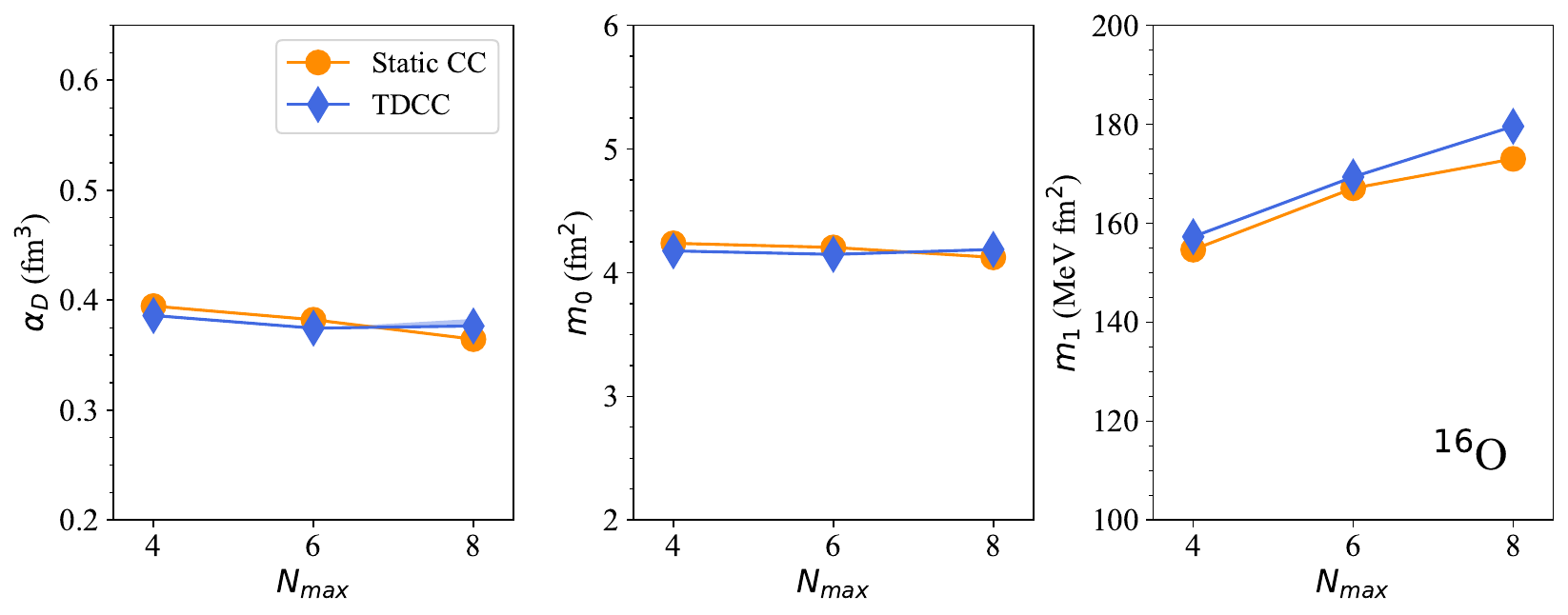}
    \caption{Comparison between the static and time-dependent results for the dipole polarizability $\alpha_D$, the non-energy-weighted sum rule $m_0$ and the energy-weighted sum rule $m_1$ in $^{16}$O for different model-space sizes. }
    \label{fig: sr-bench-16o}
\end{figure*}

Let us now compare the static and time-dependent approaches on dipole sum rules. Figs.~\ref{fig: sr-bench-4he} and~\ref{fig: sr-bench-16o} show results for $\alpha_D$, $m_0$ and $m_1$ as a function of $N_{\rm max}$ for $^{4}$He and $^{16}$O respectively. 

In the static framework, moments of the response function are calculated from the LIT, taking the limit for $\Gamma\rightarrow 0$~\cite{miorelli2016}. The results in  Figs.~\ref{fig: sr-bench-4he} and~\ref{fig: sr-bench-16o} have been obtained with $\Gamma = 10^{-4}$ MeV, which is sufficient to converge the integral. While in the static calculation we can easily make the integration grid denser, in the time-dependent one we are limited by the energy resolution of the response, which in this case corresponds to about $0.6$ MeV. For this reason, we performed the integration employing both a linear and a cubic-spline interpolator. In most cases, differences between the two approaches remain below the percent level. Only for $^{16}$O and $N_{\rm max} = 8$, the results show a larger deviation, reaching around $1\%$ for $\alpha_D$. We attribute this to the higher degree of fragmentation of the strength around the main peak, as seen in Fig.~\ref{fig:response-conv-16O}. To emphasize these cases, Figs.~\ref{fig: sr-bench-4he} and~\ref{fig: sr-bench-16o} show the results obtained with cubic splines, together with a band representing the difference between the linear and cubic-spline interpolations. In most instances, this band is too narrow to be visible. 

Figures~\ref{fig: sr-bench-4he} and~\ref{fig: sr-bench-16o} show a good agreement between the static and time-dependent results for sum rules. 
For both nuclei and for every value of $N_{\rm max}$, the central values for the three sum rules in the static and time-dependent approaches deviate on average by around $2\%$, and in general by no more than $4\%$. The largest deviations are observed in the case of the energy-weighted sum rule $m_1$. Clearly, $m_1$ is mostly affected by the high-energy part of the spectrum which is cut off beyond about $200$~MeV because of the time step $\Delta t$ employed in TDCC simulations. For example, if we limit the integration to below 200~MeV, the difference between $m_1$ from the static and TDCC response is reduced from $4\%$ to $1.2\%$ in $^{16}$O for $N_{\rm max} = 8$. 

We can compare our results for $^4$He to Ref.~\cite{baker2020}, where NCSM and symmetry-adapted NCSM values obtained with the NNLO$_{\rm opt}$ interaction are reported. Differences between our $N_{\rm max} = 8$ values and NCSM amount to around $1\%$ for $m_1$ and $m_0$, and to $5\%$ for $\alpha_D$. Given that our $N_{\rm max}=8$ values are not fully converged and are obtained within the CCSD approximation, the two approaches appear consistent with each other. Reference~\cite{miorelli2018} reported (albeit for a different two-body interaction) that the effect of triples correlations on $\alpha_D$ and $m_0$ in $^4$He is about $2\%$. This is in line with the level of agreement observed here.

\begin{figure*}[htb]
    \centering
    \includegraphics[width=0.95\textwidth]{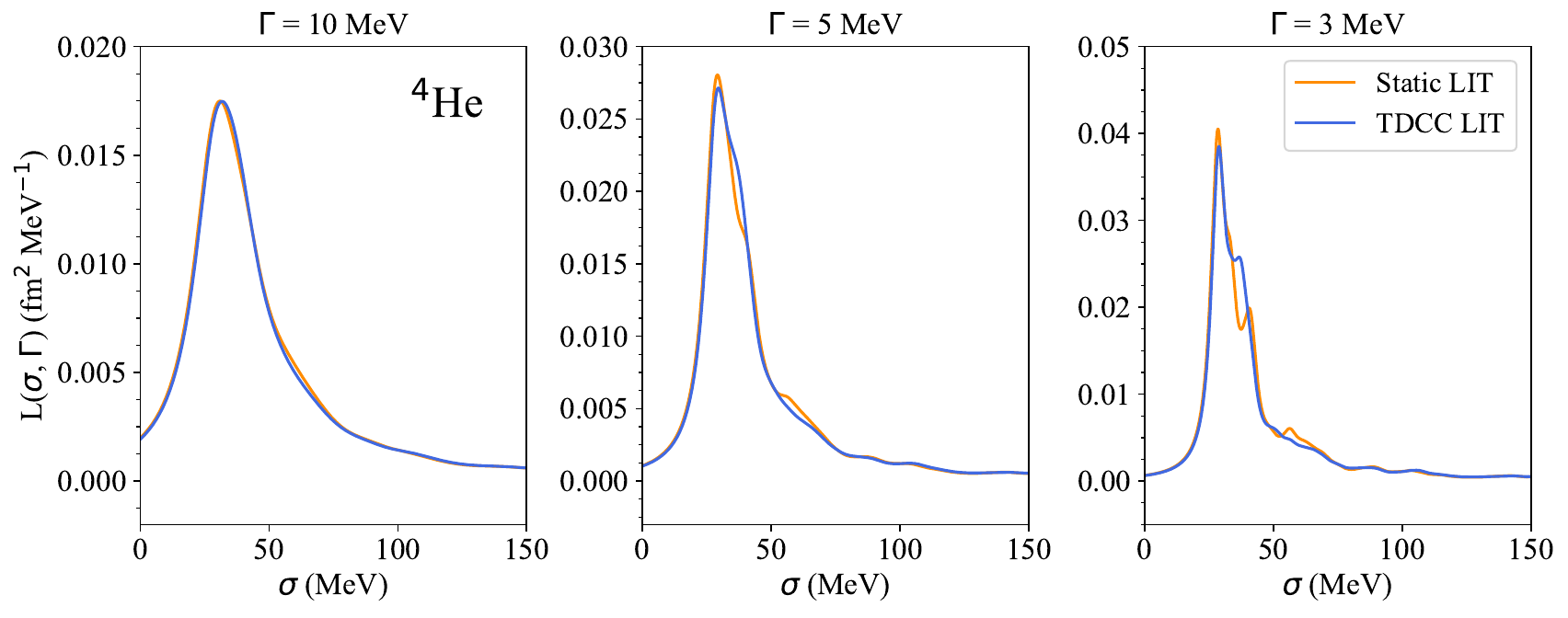}
    \caption{Comparison between the static and time-dependent results for the LIT of the response function with $\Gamma = 3, 5, 10$ MeV and $N_{\rm max} = 8$ in $^4$He. }
    \label{fig:lit-4he}
\end{figure*}

\begin{figure*}[htb]
    \centering
    \includegraphics[width=0.95\textwidth]{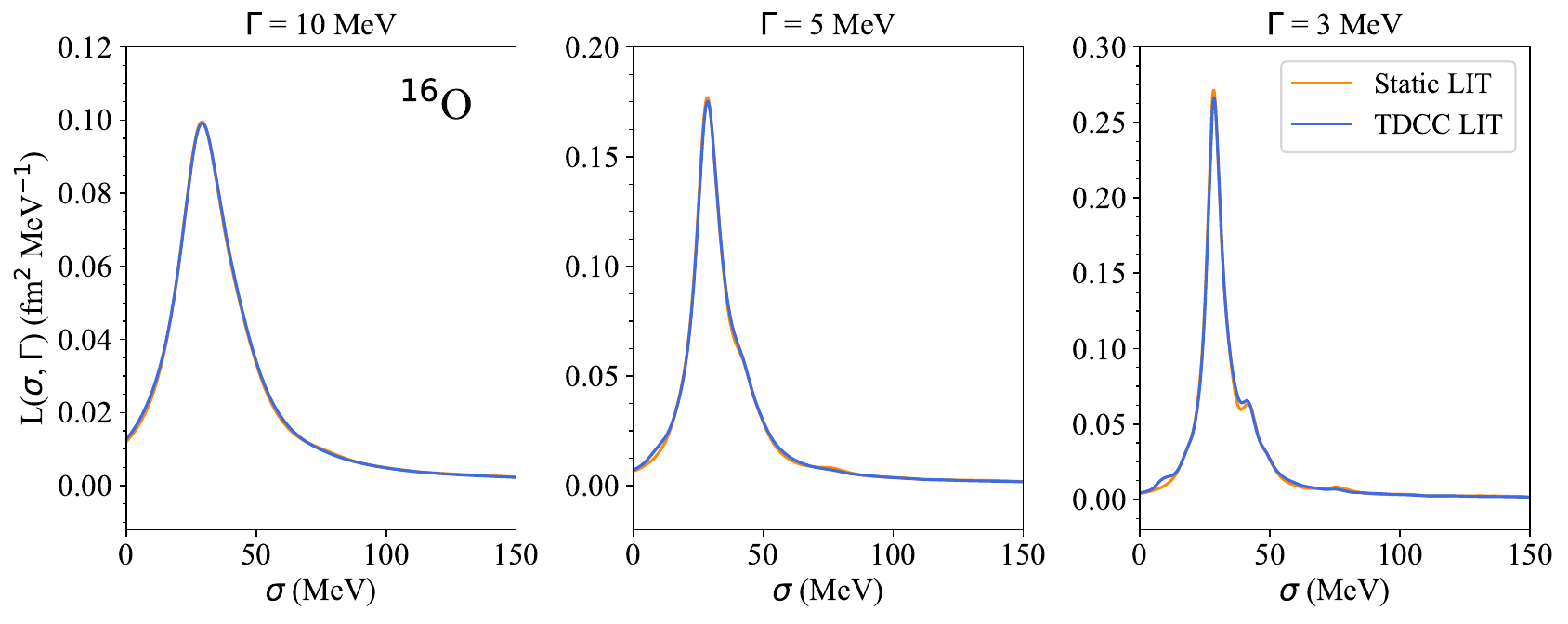}
    \caption{Comparison between the static and time-dependent results for the LIT of the response function with $\Gamma = 3, 5, 10$ MeV and $N_{\rm max} = 8$ in $^{16}$O.}
    \label{fig:lit-16o}
\end{figure*}

Let us now consider the LITs of static and time-dependent response functions. 
In Figs.~\ref{fig:lit-4he} and~\ref{fig:lit-16o} we show LIT results from the static and TDCC approaches, obtained with $\Gamma = 3, 5, 10$~MeV, for $^4$He and $^{16}$O, respectively. Overall, the LITs obtained from time-dependence are close to the static result for both nuclei. In both cases, deviations between the LIT curves start appearing by progressively decreasing the value of $\Gamma$. This effect becomes particularly evident in the case of $^4$He, whose response is characterized by few excited states at energies above the GDR peak, as shown in Fig.~\ref{fig:response-conv-4He}. Since the choice of $\Gamma$ determines the amount of smearing introduced in the response, reducing $\Gamma$ makes it easier to detect differences between the static and TDCC results. This is because the time-dependent response is characterized by a finite resolution~(\ref{resolution}). As $\Gamma$ approaches $\Delta E$, discrepancies between the two approaches become more apparent.

\subsection{Convergence in $T_{\rm max}$}
\label{sec: tmax-conv}
\begin{figure}[htb]
    \centering
    \includegraphics[width=0.49\textwidth]{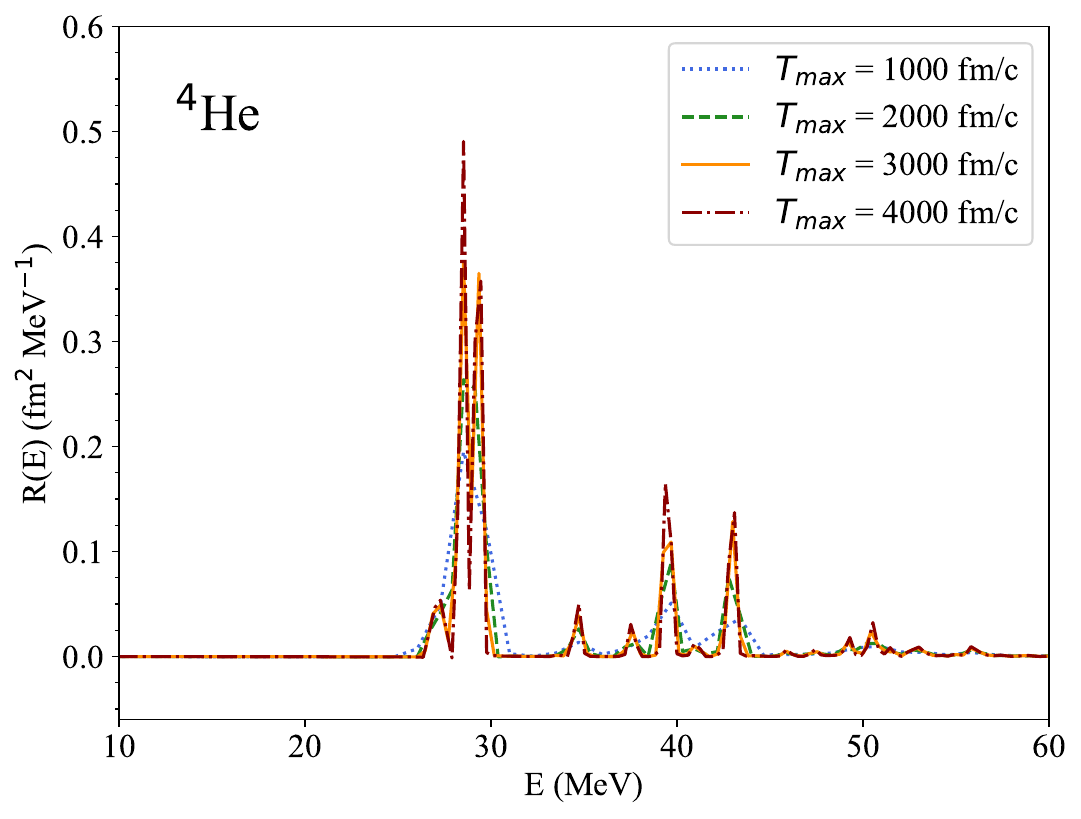}
    \caption{Response function of $^{4}$He at $N_{\rm max} = 6$ for different values of the maximum simulation time $T_{\rm max}$.}
    \label{fig: tmax-conv}
\end{figure}

In this Section, we assess the convergence of our results with respect to the maximum simulation time $T_{\rm max}$ reached in our calculations. We focus on the case of $^{4}$He and adopt a model space size of $N_{\rm max} = 6$, with which one can easily run simulations lasting over $4000$~fm/$c$. In Fig.~\ref{fig: tmax-conv} we show the dipole response of $^{4}$He obtained with $T_{\rm max}$ ranging from $1000$ to $4000$~fm/$c$. This corresponds to an energy resolution varying between $1.2$ and $0.3$ MeV. 

While the response at $T_{\rm max}=1000$~fm/$c$ is not fully resolved, particularly at higher excitation energies, the peak positions become stable at $T_{\rm max}=2000$~fm/$c$. The values of the sum rules vary by less than $0.05\%$ when increasing $T_{\rm max}$ from $1000$ to $4000$~fm/$c$. This suggests that $T_{\rm max}=2000$~fm/$c$ provides a reasonable compromise between computational cost and the accuracy of the results.

\begin{figure}[htb]
    \centering
    \includegraphics[width=0.49\textwidth]{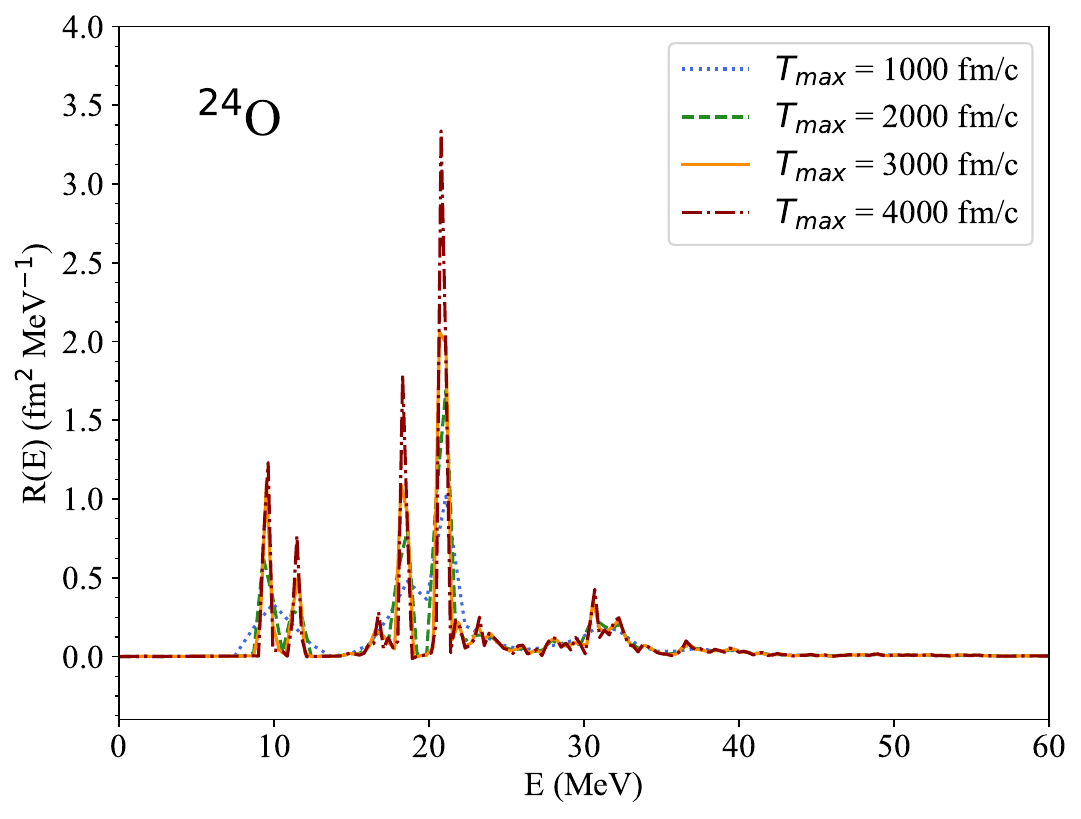}
    \caption{Response function of $^{24}$O at $N_{\rm max} = 6$ for different values of the maximum simulation time $T_{\rm max}$.}
    \label{fig:tmax-conv-24o}
\end{figure}

One could argue that longer simulation times might be needed to resolve low-energy modes, e.g. when considering the dipole response of neutron-rich nuclei, characterized by a strength enhancement at low energy, the so-called pygmy dipole resonance (PDR)~\cite{aumann2013,lanza2023}. For this reason, we study the convergence in $T_{\rm max}$ for the dipole response function of $^{24}$O and show results in Fig.~\ref{fig:tmax-conv-24o}.

The response function shows a pronounced fragmentation in the $10$–$40$~MeV region, with low-lying peaks emerging around $10$~MeV. The GDR is located near $20$~MeV. Also in this case, a simulation time of $T_{\rm max}=2000$~fm/$c$ is sufficient to stabilize the location of the peaks. Although our results are obtained using only a two-nucleon interaction, we can compare it to the one  reported in Ref.~\cite{bonaiti2024}, which employs the $\Delta$NNLO$_{\rm GO}$(394) interaction including three-nucleon potentials. In that work, while the first low-lying dipole state is found at lower energies, around $5$~MeV, the GDR is again located at $20$~MeV, consistent with our result.  Without accounting for convergence uncertainties, we find that the TDCC results for $\alpha_D$ and $m_0$ are about $30\%$ smaller than the CCSD values of Ref.~\cite{bonaiti2024}. We attribute this discrepancy to a lack of dipole strength in the absence of three-nucleon forces. 

\subsection{Density fluctuations}
In the previous Sections, we demonstrated the equivalence between static and time-dependent approaches to response functions, providing a benchmark in the case of $^4$He and $^{16}$O. We now turn to features that are unique to a time-dependent calculation and can complement the information obtained from a static framework. While static approaches are computationally cheaper, a time-dependent calculation gives direct access to dynamical quantities that cannot be easily inferred otherwise. 

For instance, in TDCC simulations we can study the evolution in time of the one-body density matrix of Eq.~(\ref{eq:density-mat}). From the latter, one can derive the nuclear matter, point-proton and point-neutron radial densities. Here, we focus on the point-proton and point-neutron densities. It is particularly instructive to analyze the behavior of $\rho(r,t)-\rho(r,t=0)$, corresponding to the fluctuation of the nuclear density with respect to its initial condition at $t=0$. In our case, the latter is given by the nuclear ground state. Density fluctuations reveal how the spatial distribution of nucleons evolves in time, and they can be efficiently visualized in movies (available in the Supplemental Material~\cite{supplementalmaterial}). Here, we focus on representative snapshots at fixed time $t=500$~fm/$c$.

\begin{figure}[htb]
    \centering
    \includegraphics[width=0.49\textwidth]{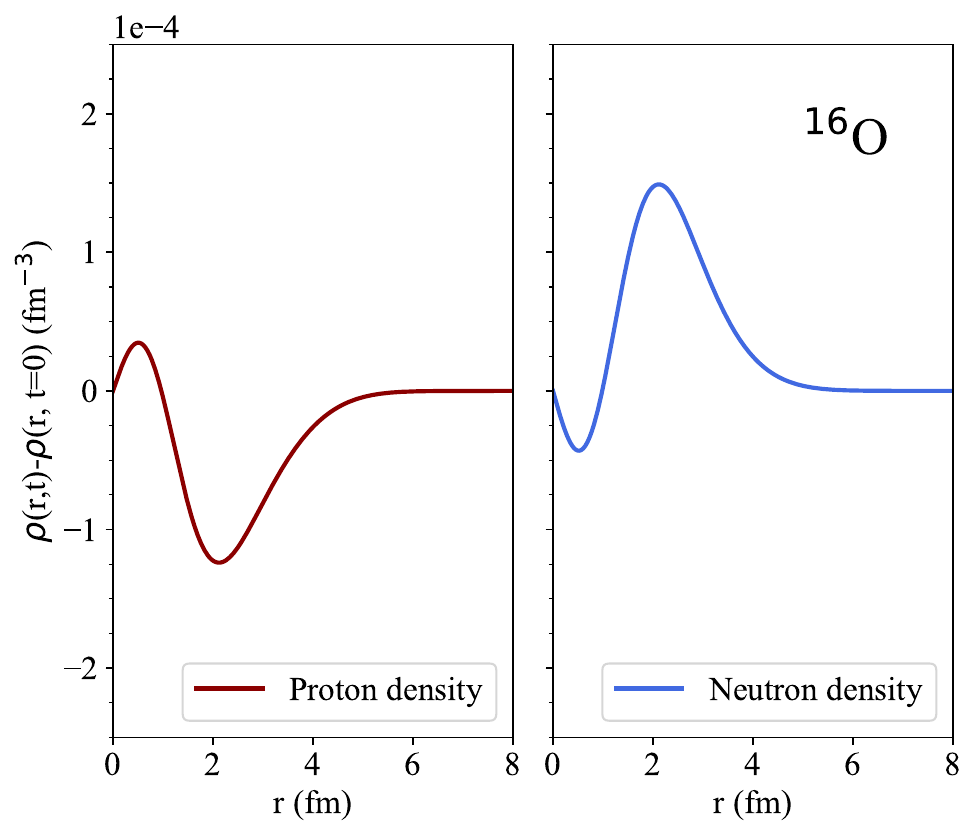}
    \caption{Snapshot at $t = 500$~fm/$c$ of the point-proton (left panel) and point-neutron (right panel) density fluctuations for $^{16}$O. }
    \label{fig:dens-fluc-16o}
\end{figure}

First, we focus on the case of $^{16}$O. The response of $^{16}$O is dominated by the GDR mode, as shown in Fig.~\ref{fig:response-conv-16O}. The corresponding density fluctuations at $t=500$~fm/$c$ are displayed in Fig.~\ref{fig:dens-fluc-16o}. 

The oscillations of proton and neutron densities are periodic and occur in counterphase, providing a direct visualization of the GDR as a collective motion of protons against neutrons in the nucleus~\cite{goldhaber1948,steinwedel1950}.

We can also go a step further. Neutron-rich nuclei are characterized by the low-lying PDR mode, often interpreted as a vibration of the excess neutrons against a core~\cite{aumann2013}, and correlated with the development of a neutron skin~\cite{in2025}. Such low-energy strength emerges for example in the response of $^{24}$O, recall  Fig.~\ref{fig:tmax-conv-24o}. 

\begin{figure}[htb]
    \centering
    \includegraphics[width=0.49\textwidth]{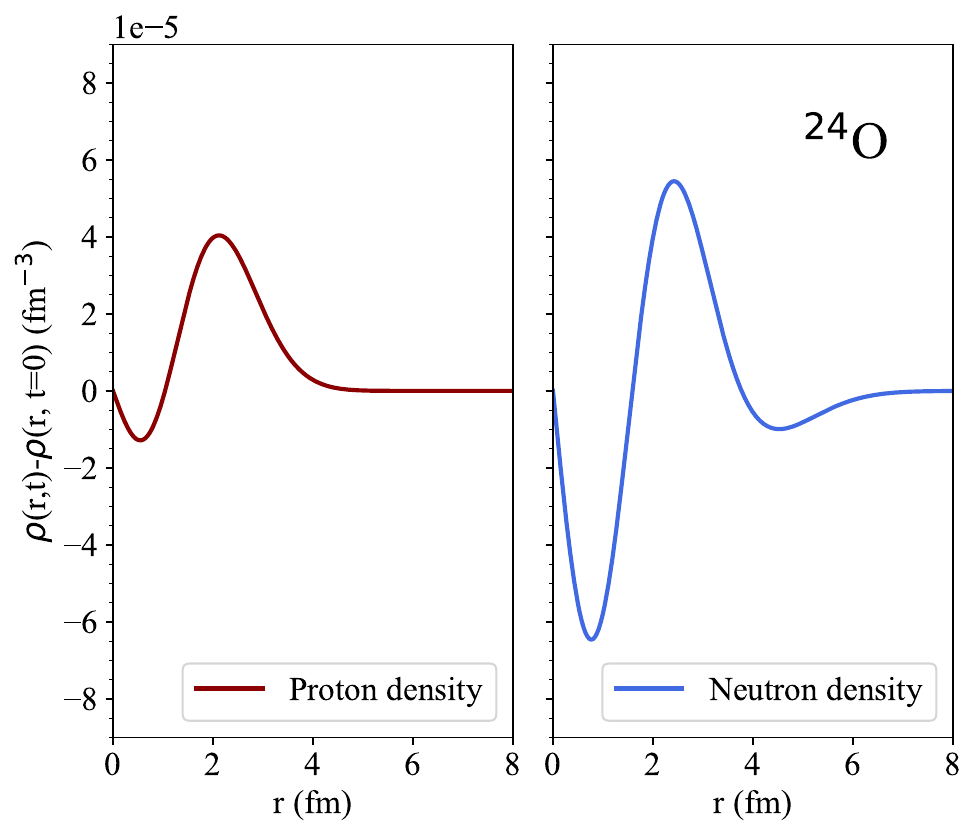}
    \caption{Snapshot at $t = 500$~fm/$c$ of the point-proton (left panel) and point-neutron (right panel) density fluctuations corresponding to the lowest energy dipole mode in the response of $^{24}$O. }
    \label{fig:dens-fluc-24o}
\end{figure}

We now isolate the contribution of the PDR to the density fluctuations in $^{24}$O. To this aim, we analyze the time-dependent densities as follows. Starting from $\rho(r,t)$ for $N_{\max}=6$, we perform a Fourier transform along the time axis and apply a box filter to select the frequency window corresponding to the first excited state ($8$–$10.5$ MeV). An inverse Fourier transform allows to retrieve the corresponding density oscillation. Such a procedure removes contributions from higher-lying excitations and reconstructs the density fluctuation associated with the PDR alone. The resulting density fluctuations at $t=500$~fm/$c$ are shown in Fig.~\ref{fig:dens-fluc-24o}.

For radii up to about $4$~fm, proton and neutron density fluctuations oscillate in phase. Beyond $4$~fm, the proton fluctuation essentially vanishes, while the neutron density develops an additional oscillatory component at the nuclear surface. Such an oscillation appears in counterphase with respect to the core and represents a signature of the presence of a PDR mode for this nucleus. 

Figures~\ref{fig:dens-fluc-16o} and~\ref{fig:dens-fluc-24o} show a key advantage of a time-dependent approach to responses. The latter makes a real-time observation and interpretation of the giant and pygmy dipole modes possible, a feature  which is hardly accessible in static calculations.

\subsection{Spectral information in the non-linear regime}
\label{sec: nonlinear}
Up to now, we have worked within perturbation theory, assuming that the external electric dipole field acting on the nucleus is characterized by a small field intensity $\varepsilon$. In this situation, we are in the so-called linear regime, where the time-dependent transition moment $D(t)$ is directly proportional to $\varepsilon$. The  time-dependent description also allows us to explore the regime of non-linear motion, emerging for a field intensity $\varepsilon\gg 1$ MeV/fm. This is hardly achievable in a static approach. 

\begin{figure}[htb]
    \centering
    \includegraphics[width=0.49\textwidth]{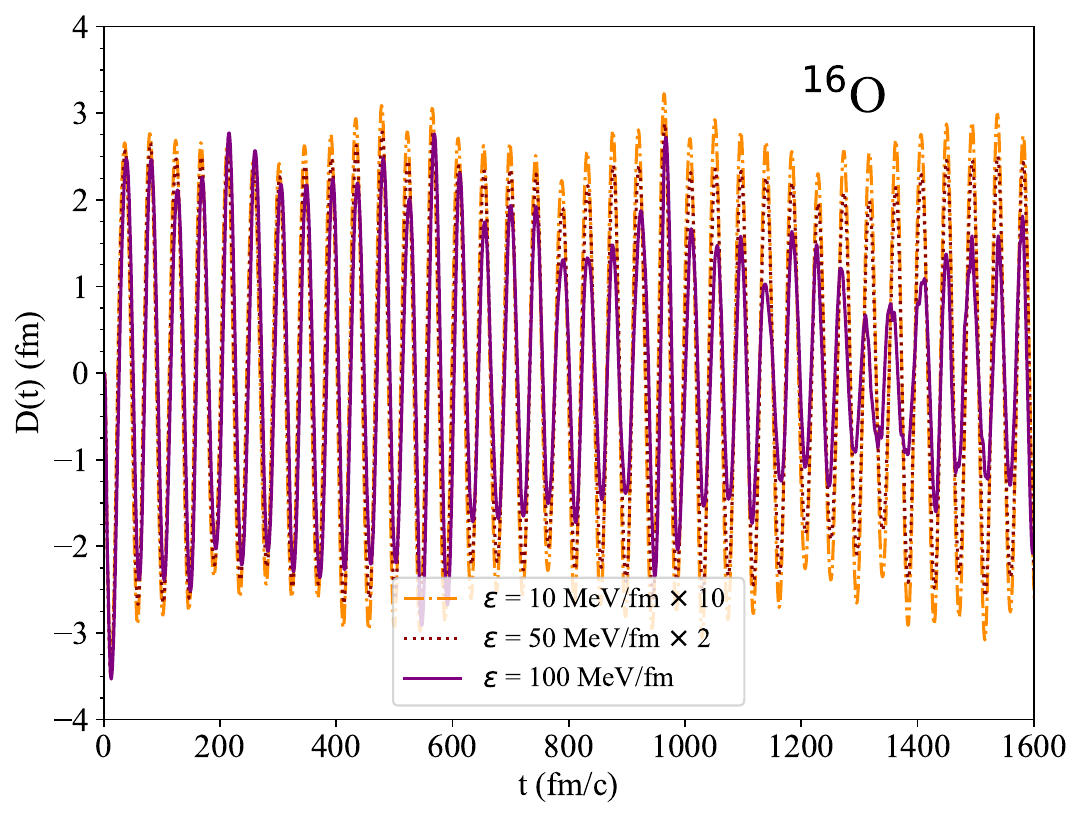}
    \caption{Time-dependent electric dipole moment of $^{16}$O for $\varepsilon = 10, 50, 100$ MeV/fm and a model-space size of $N_{\rm max} = 6$. The curves corresponding to $\varepsilon = 10, 50$ MeV/fm have been multiplied by a factor $10$ and $2$, respectively. See details in the text.}
    \label{fig: nl-dipole}
\end{figure}

It is interesting to analyze the energy scales at which such non-linear motion emerges. Let us focus on the case of $^{16}$O. We can assume non-linearities to appear when the magnitude of the external field $\varepsilon D$ applied to the nucleus becomes comparable to typical nuclear excitation energies, around $1$~MeV. As a rough estimate of the induced electric dipole moment, we can use the square root of the transition strength $B(E1, 1_1^-\rightarrow 0^+)$ between the first dipole-excited state and the ground state in $^{16}$O. Experimentally, $B(E1, 1_1^-\rightarrow 0^+) = 1.43(8)\times 10^{-4}\; e^2$fm$^2$~\cite{tilley1993}. Therefore, $\sqrt{B(E1)} \approx 0.01\; e\;$fm. This suggests that a field strength of order $100$ MeV/fm is needed to yield sizable non-linear effects. 
In Fig.~\ref{fig: nl-dipole}, we show the time-dependent dipole moment of $^{16}$O for $\varepsilon = 10, 50, 100$~MeV/fm and a model-space size of $N_{\rm max} = 6$. 

We have verified that with $\varepsilon = 10$ MeV/fm we are safely in the linear regime. In Fig.~\ref{fig: nl-dipole}, the curves obtained for $\varepsilon = 10, 50$~MeV/fm have been multiplied by a factor $10$ and $2$, respectively, so that they would coincide with the $\varepsilon = 100$~MeV/fm result if the system was in the linear regime. It is clear that non-linear motion emerges already starting from $\varepsilon = 50$~MeV/fm, and even more clearly for $\varepsilon = 100$~MeV/fm, in agreement with our naive estimate. Moreover, while the periodic oscillations of $D(t)$ at $\varepsilon = 50$~MeV/fm still follow closely those obtained for $\varepsilon = 10$~MeV/fm (albeit a reduced amplitude) more sizable deviations, indicating the emergence of additional frequency modes, are observed for $\varepsilon = 100$~MeV/fm. 

\begin{figure}[htb]
    \centering
    \includegraphics[width=0.49\textwidth]{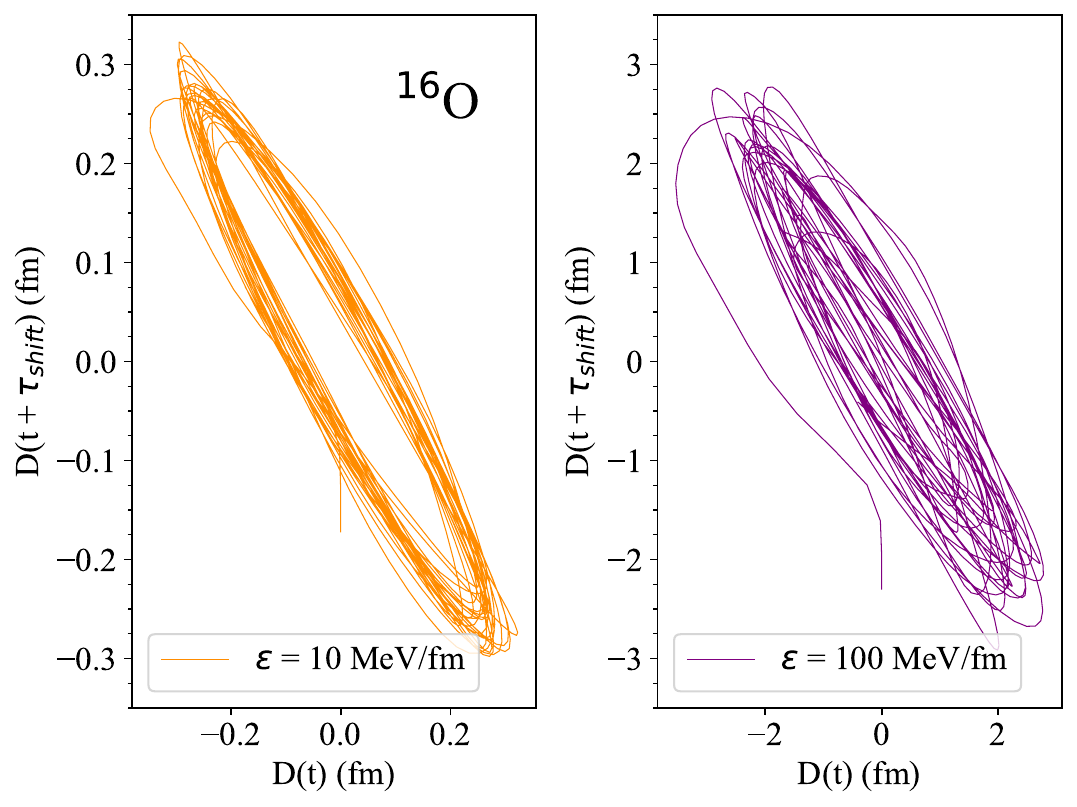}
    \caption{Time-dependent dipole moment $D(t)$ of $^{16}$O at $N_{\rm max} = 6$ as a function of its corresponding shifted signal $D(t+\tau_{\rm shift})$ with $\tau_{\rm shift} = 20$~fm/$c$ for $\varepsilon = 0.1$~MeV/fm (left panel) and $\varepsilon = 100$~MeV/fm (right panel). Time trajectories are plotted up to $t = 1000$~fm/$c$.}
    \label{fig: delayed-dt}
\end{figure}

Let us focus on the case of a field strength $\varepsilon = 100$~MeV/fm. It is instructive to study the nature of the non-linear effects in Fig.~\ref{fig: nl-dipole}. In the regime of non-linear dynamics, in fact, chaotic patterns can emerge, as shown in the case of isovector monopole excitation in Ref.~\cite{vretenar1997}. Following that work, a simple way to detect chaotic patterns is to plot the time evolution of $D(t)$ versus itself, but delayed by a fixed time constant $\tau_{\rm shift}$, which only needs to be different from a natural period of the system. If the time evolution of the system is chaotic, the corresponding trajectories in the plane $[D(t),\;D(t+\tau_{\rm shift})]$ do not close. In Fig.~\ref{fig: delayed-dt}, we plot $D(t)$ for $^{16}$O as a function of the corresponding $D(t+\tau_{\rm shift})$ with $\tau_{\rm shift} = 20$~fm/$c$, in the linear ($\varepsilon = 10$~MeV/fm) and non-linear ($\varepsilon = 100$~MeV/fm) regime.

The evolution of the time trajectories in the linear regime (left panel) outlines closed ellipses, indicating a regular motion. On the other hand, for the non-linear signal (right panel) the phase space trajectories suggest the emergence of chaotic behavior in this strong field limit. It is interesting to notice that clear chaotic patterns in time trajectories begin to emerge in the non-linear regime at relatively long times, between $600$ and $1000$~fm/$c$, suggesting that a relative large $T_{\rm max}$ is needed to resolve these effects. 

\begin{figure}[htb]
    \centering
    \includegraphics[width=0.49\textwidth]{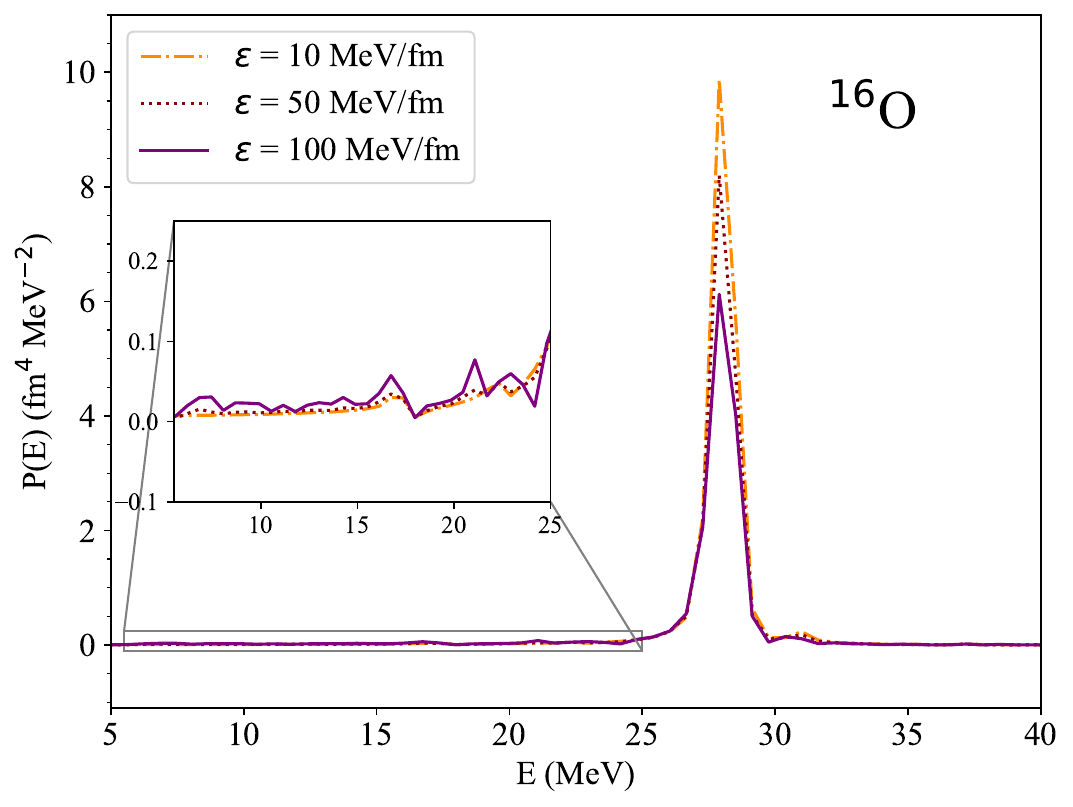}
    \caption{Power spectrum of $^{16}$O for $\varepsilon = 10, 50, 100$~MeV/fm and $N_{\rm max} = 6$.}
    \label{fig: power-spectrum}
\end{figure}

It is instructive to analyze the effects of non-linearities in the Fourier domain. This also allows for a qualitative comparison with previous TDDFT results~\cite{reinhard2007}. In the non-linear regime, we are considering the behavior of the nucleus in a strong electric field, outside of the range of applicability of perturbation theory, on which the derivation of Section~\ref{sec: pert-theory} is based. In this regime, spectral information is often analyzed by considering the so-called power spectrum~\cite{calvayrac1997}
\begin{equation}
    P(E) = \left|\frac{\widetilde{D}(E)}{\varepsilon \widetilde{g}(E)}\right|^2. 
\end{equation}
In Fig.~\ref{fig: power-spectrum} we plot $P(E)$ for different values of $\varepsilon$, employing $T_{\rm max} = 2000$~fm/$c$. 

Despite the chaotic behavior emerging from the dipole transition moment, the strength remains mostly concentrated in the GDR peak at around 28~MeV. However, we observe a significant $40\%$ reduction in the height of the GDR with increasing $\varepsilon$. Also, as shown by the inset of Fig.~\ref{fig: power-spectrum}, the spectrum appears to be more fragmented and characterized by emerging low-lying strength when going towards the non-linear regime. Qualitatively, these small anharmonicities are consistent with the TDDFT results of Ref.~\cite{reinhard2007}, which reported a reduction in the GDR peak intensity and an enhancement of the response at low-energy for $^{16}$O. 

One might argue that such strong fields are purely of theoretical interest and beyond experimental reach. However, the proposed Gamma Factory at CERN, currently under consideration within the Physics Beyond Colliders program, would in principle allow for the production of photons with energies up to $400$~MeV~\cite{budker2022}. At this energy, the photon wavelength becomes of order $1$~fm, comparable to typical nuclear length scales. This suggests that the regime we consider here, where the electric field becomes comparable with typical nuclear excitations, might become accessible to experimental investigation in the future. 

\section{Conclusions}
\label{sec: conclusion}
We used time-dependent coupled-cluster theory to compute nuclear response functions.
For validation, we focused on electric dipole transitions in $^{4}$He and $^{16}$O. When considering moments of the response function distribution ($\alpha_D$, $m_0$, and $m_1$), we observed negligible discrepancies between TDCC and static LIT-CC results when accounting for the limited resolution of the TDCC approach. Our findings are also consistent with previous NCSM investigations. Moreover, we observed that a maximum simulation time of $2000$~fm/$c$, corresponding to an energy resolution of $0.3$~MeV, is sufficient to resolve the main features of the strength distribution, even in neutron-rich nuclei such as $^{24}$O, where the pygmy dipole mode emerges.

We showed that a time-dependent description of nuclear responses gives access to information that is difficult to obtain in a static framework. Analysis of the time evolution of proton and neutron densities relative to the nuclear ground state provides visual evidence of the pygmy and giant dipole resonances as collective motion of protons against neutrons in $^{16}$O and $^{24}$O. Movies following the evolution of density fluctuations in time are included in the Supplemental Material. This method also enables to investigate the behavior of the nucleus in strong electric fields. In that regime, our results suggest the emergence of chaotic dynamics. The spectral information obtained in this regime qualitatively agrees with previous time-dependent mean-field results, indicating slight anharmonicities, as a lowering of the GDR peak. 

Our current implementation, based on a Hartree–Fock single-particle basis, does not adequately treat continuum degrees of freedom, limiting the description of phenomena such as damping widths of giant resonances and particle emission~\cite{stringari1979,chomaz1987,avez2013}. These limitations could be overcome through a lattice formulation of coupled-cluster theory,  which has recently been developed~\cite{rothman2025a,rothman2025b}, in combination with absorbing boundary conditions or with the use of large lattices~\cite{reinhard2006,avez2013}. A lattice-based approach would also facilitate the extension of this framework to nuclear collisions. In this regard, following previous work in quantum chemistry~\cite{kvaal2012,sato2018,kristiansen2022}, we also plan to study the effect of time dependence in the single-particle orbitals, a feature which could be helpful in an accurate description of reaction processes. 

\section*{Data availability}
The data that support the findings of this article are openly available~\cite{bonaiti_tdcc_zenodo2025}.

\begin{acknowledgments}
F.B. thanks Gustav R. Jansen for useful discussions and suggestions on how to make TDCC runs more efficient, and Sonia Bacca and Aaina Thapa for useful discussions. This work was supported by the U.S. Department of Energy, Office of Science, Office of Nuclear Physics, under the FRIB
Theory Alliance award DE-SC0013617 and  Award Nos.~DE-FG02-96ER40963; by the U.S. Department of Energy under the Award No. DOE-DE-NA0004074 (NNSA, the Stewardship Science Academic Alliances program); by the U.S. Department of Energy, Office of Science, Office of Advanced Scientific Computing Research and Office of Nuclear Physics, Scientific Discovery through Advanced Computing (SciDAC) program (SciDAC-5 NUCLEI); by the Alexander von~Humboldt Foundation. This work was funded in part by the U.S. Department of Energy, Office of Science,
Office of Advanced Scientific Computing Research, Scientific Discovery through Advanced
Computing (SciDAC) Program through the FASTMath Institute. This research used resources of the Oak Ridge Leadership Computing Facility located at Oak Ridge National Laboratory, which is supported by the Office of Science of the Department of Energy under contract No. DE-AC05-00OR22725. Computer time was provided by the Innovative and Novel Computational Impact on Theory and Experiment (INCITE) program, by the Institute for Cyber-Enabled Research at Michigan State University and by the supercomputer Mogon at Johannes Gutenberg Universit\"at Mainz.
\end{acknowledgments}

\bibliography{master}
\end{document}